\documentclass[journal=jpcafh,manuscript=article]{achemso}
\usepackage[version=4]{mhchem}
\usepackage{commath}
\usepackage{subcaption}
\usepackage{multirow}
\usepackage{booktabs}
\usepackage{placeins}

\author{Jiachen Li}
\affiliation{Department of Chemistry, Duke University, Durham, NC 27708, USA}
\author{Jincheng Yu}
\affiliation{Department of Chemistry, Duke University, Durham, NC 27708, USA}
\author{Zehua Chen}
\affiliation{Department of Chemistry, Duke University, Durham, NC 27708, USA}
\author{Weitao Yang}
\affiliation{Department of Chemistry, Duke University, Durham, NC 27708, USA}
\email{weitao.yang@duke.edu}

\title{Linear Scaling Calculations of Excitation Energies with Active-Space Particle-Particle Random Phase Approximation}

\begin{document}

\begin{abstract}
We developed an efficient active-space particle-particle random phase approximation (ppRPA) approach to calculate accurate charge-neutral excitation energies of molecular systems.
The active-space ppRPA approach constrains both indexes in particle and hole pairs in the ppRPA matrix,
which only selects frontier orbitals with dominant contributions to low-lying excitation energies.
It employs the truncation in both orbital indexes in the particle-particle and the hole-hole spaces.
The resulting matrix, 
the eigenvalues of which are excitation energies, 
has a dimension that is independent of the size of the systems.
The computational effort for the excitation energy calculation, 
therefore, 
scales linearly with system size and is negligible compared with the ground-state calculation of the ($N-2$)-electron system, 
where $N$ is the electron number of the molecule. 
With the active space consisting of $30$ occupied and $30$ virtual orbitals, 
the active-space ppRPA approach predicts excitation energies of valence, 
charge-transfer, Rydberg, double and diradical excitations
with the mean absolute errors (MAEs) smaller than $0.03$ \,{eV} compared with the full-space ppRPA results.
As a side product, 
we also applied the active-space ppRPA approach in the renormalized singles (RS) T-matrix approach. 
Combining the non-interacting pair approximation that approximates the contribution to the self-energy outside the active space,
the active-space $G_{\text{RS}}T_{\text{RS}}$@PBE approach predicts accurate absolute and relative core-level binding energies with the MAE around $1.58$ \,{eV} and $0.3$ \,{eV}, respectively.
The developed linear scaling calculation of excitation energies is promising for applications to large and complex systems.
\end{abstract}

\section{INTRODUCTION}
The accurate description for charge-neutral (optical) excitations is important for understanding optical spectroscopies.
In past decades,
many theoretical approaches have been developed to compute excitation energies.
Time-dependent density functional theory\cite{casidaTimeDependentDensityFunctional1995,ullrichTimeDependentDensityFunctionalTheory2011,rungeDensityFunctionalTheoryTimeDependent1984} (TDDFT) ranks among the most used approach  for finite systems because of the good compromise between the accuracy and the computational cost.
TDDFT with commonly used density functional approximations (DFAs) has been widely applied to predict excitation energies of different systems including molecules, liquids and solids\cite{casidaLinearResponseTimeDependentDensity2006,casidaTimedependentDensityfunctionalTheory2009,laurentTDDFTBenchmarksReview2013}.
However, it is well-known that TDDFT with conventional DFAs provides an incorrect long-range behavior\cite{tozerRelationshipLongrangeChargetransfer2003,dreuwLongrangeChargetransferExcited2003}.
Consequently, TDDFT fails to describe Rydberg and charge-transfer (CT) excitations\cite{tozerRelationshipLongrangeChargetransfer2003,dreuwLongrangeChargetransferExcited2003}.
Efforts including using range-separated functionals\cite{leiningerCombiningLongrangeConfiguration1997,besleyTimedependentDensityFunctional2009,steinPredictionChargetransferExcitations2009,refaely-abramsonFundamentalExcitationGaps2011} and tuning the amount of the Hartree-Fock (HF) exchange in DFAs\cite{brucknerBenchmarkingSingletTriplet2017,gongBenchmarkParameterTuning2020,kempfer-robertsonRoleExactExchange2022} are made to address this issue.
In addition,
the accuracy of TDDFT largely depends on the exchange-correlation (XC) kernel\cite{laurentTDDFTBenchmarksReview2013}.
Different from TDDFT,
the Bethe-Salpeter equation\cite{shamManyParticleDerivationEffectiveMass1966,hankeManyParticleEffectsOptical1979,salpeterRelativisticEquationBoundState1951} (BSE) formalism has also gained increasing attentions.
BSE commonly takes the quasiparticle (QP) energies from the $GW$ approximation\cite{golzeGWCompendiumPractical2019,martinInteractingElectrons2016,reiningGWApproximationContent2018,hedinNewMethodCalculating1965} to calculate the excitation energies,
which is the BSE/$GW$ approach.
Because of the correct description for the long-range behavior and the dynamical screening interaction in real systems,
BSE/$GW$ is shown to predict accurate excitation energies for various kinds of systems\cite{yaoAllElectronBSEGW2022,mckeonOptimallyTunedRangeseparated2022,knyshModellingExcitedState2022,balasubramaniStaticPolarizabilitiesGeneralized2022,liCombiningRenormalizedSingles2022}.
However,
the accuracy of the BSE/$GW$ formalism strongly depends on the level of the self-consistency in the preceding $GW$ calculation.
BSE combined with the one-shot $G_0W_0$ method has the undesired starting point dependence\cite{maromBenchmarkGWMethods2012,ducheminCubicScalingAllElectronGW2021,liCombiningRenormalizedSingles2022},
similar to TDDFT.
BSE combined with the eigenvalue-self-consistent $GW$ (ev$GW$) method shows great improvements over BSE/$G_0W_0$ but brings extra computational costs\cite{forsterQuasiparticleSelfConsistentGWBethe2022,jacqueminBenchmarkBetheSalpeterTriplet2017,liCombiningRenormalizedSingles2022}.
Efforts including combining BSE with the renormalized singles Green's function\cite{liCombiningRenormalizedSingles2022} and using the optimal starting point\cite{mckeonOptimallyTunedRangeseparated2022} have been made to improve the accuracy and reduce the computational cost.
Recently, approaches combining BSE with generalized Kohn-Sham (KS) methods, 
including localized orbital scaling correction\cite{liCombiningLocalizedOrbital2022} (LOSC) and Koopmans-compliant functionals\cite{elliottKoopmansMeetsBethe2019},
have been developed to bypass the computationally demanding $GW$ calculation.
BSE combined with the T-matrix approximation has also been reported\cite{loosStaticDynamicBethe2022}. 
In addition to the computationally affordable BSE and TDDFT formalisms,
wave function approaches with high accuracy are also used to predict excitation energies, 
including complete active space second-order perturbation theory\cite{anderssonSecondOrderPerturbation1992} (CASPT2) and coupled cluster methods\cite{christiansenResponseFunctionsCC31995,kochExcitationEnergiesCoupled1990} (CC3, CCSDT). 
However, due to the demanding computational cost, 
wave function approaches are usually used for benchmarks\cite{loosMountaineeringStrategyExcited2018,loosMountaineeringStrategyExcited2020,verilQUESTDBDatabaseHighly2021}.

The particle-particle random phase approximation (ppRPA), originally developed to describe nuclear many-body correlation\cite{ringNuclearManybodyProblem1980,ripkaQuantumTheoryFinite1986,yangDoubleRydbergCharge2013} has been applied to describe ground-state and excited-state properties of molecular systems\cite{vanaggelenExchangecorrelationEnergyPairing2013,vanaggelenExchangecorrelationEnergyPairing2014}.
The ppRPA can be derived from different approaches,
including the adiabatic connection\cite{vanaggelenExchangecorrelationEnergyPairing2013,vanaggelenExchangecorrelationEnergyPairing2014} as a parallel to the traditional adiabatic connection using the the particle-hole density fluctuations\cite{langrethExchangecorrelationEnergyMetallic1977,harrisSurfaceEnergyBounded1974}, TDDFT with the pairing field\cite{pengLinearresponseTimedependentDensityfunctional2014} and the equation of motion\cite{ringNuclearManybodyProblem1980,roweEquationsofMotionMethodExtended1968}.
As a counterpart to the commonly used particle-hole random phase approximation\cite{bohmCollectiveDescriptionElectron1951,renRandomphaseApproximationIts2012} (phRPA),
the ppRPA conveys information in the particle-particle and the hole-hole channel,
which leads to the two-electron addition and the two-electron removal energies.
Thus, the excitation energy of the $N$-electron system can be obtained by the difference between two-electron addition energies of the ($N-2$)-electron system\cite{yangDoubleRydbergCharge2013}.
The same set of equations were also applied for describing the excited states of $N$-electron molecule using the hole-hole excitations of the corresponding ($N+2$)-electron system\cite{bannwarthHoleHoleTamm2020,yuInitioNonadiabaticMolecular2020}.
The ppRPA has been shown to predict accurate excitation energies of different characters including singlet-triplet (S-T) gaps of diradicals, double excitations, CT excitations, Rydberg excitations and valence excitations.
It also describes conical intersections well\cite{yangConicalIntersectionsParticle2016}.
The success of the ppRPA for predicting excitation energies of molecular systems can be attributed to several reasons.
First,
the ppRPA can be viewed as a Fock-space embedding approach that treats two frontier electrons in a subspace configuration interaction fashion with a seamless combination of density functional theory\cite{hohenbergInhomogeneousElectronGas1964,parrDensityFunctionalTheoryAtoms1989} (DFT) for the remaining ($N-2$) electrons\cite{yangNatureGroundElectronic2016,zhangAccurateEfficientCalculation2016}.
Thus, the ppRPA provides an accurate description of the static correlation of the two non-bonding electrons in the diradical systems\cite{yangNatureGroundElectronic2016,yangSingletTripletEnergy2015}.
Second,
the ppRPA contains the information in the particle-particle channel and is naturally capable of describing double excitations\cite{yangDoubleRydbergCharge2013},
which cannot be captured by TDDFT or BSE with the adiabatic approximation.
Third,
because the ppRPA kernel provides the correct long-range asymptotic behavior,
ppRPA can accurately predict CT and Rydberg excitation energies\cite{yangChargeTransferExcitations2017,yangDoubleRydbergCharge2013}.
Finally,
the ppRPA kernel is independent of the DFA reference,
thus the starting point dependence of the ppRPA is smaller than TDDFT\cite{yangExcitationEnergiesParticleparticle2014,yangChargeTransferExcitations2017}.
Recently, the ppRPA with Tamm-Dancoff approximation (TDA) has been applied in the multireference DFT approach to provide accurate dissociation energies and excitation energies\cite{chenMultireferenceDensityFunctional2017,liMultireferenceDensityFunctional2022}.
In addition to the calculation of the excitation energy, 
the ppRPA eigenvalues and eigenvectors are also used in the T-matrix approximation to calculate QP energies\cite{zhangAccurateQuasiparticleSpectra2017,liRenormalizedSinglesGreen2021}.

However,
it is challenging to apply the ppRPA to compute excitation energies of large systems due to its relatively high computational cost.
In the ppRPA,
the two-electron integrals of four virtual orbitals are needed\cite{vanaggelenExchangecorrelationEnergyPairing2013,yangExcitationEnergiesParticleparticle2014}.
Because the number of virtual orbitals $N_{\text{vir}}$ is generally much larger than the number of occupied orbitals $N_{\text{occ}}$,
the dimension of the ppRPA is much larger than the dimension of TDDFT or BSE\cite{zhangAccurateEfficientCalculation2016}.
Although the formal scaling of solving ppRPA equation can be reduced to $\mathcal{O} (N^4)$ by using the Davidson algorithm\cite{davidsonIterativeCalculationFew1975,yangExcitationEnergiesParticleparticle2014} ($N$ is the size of the system),
the size of the ppRPA matrix has a large prefactor compared with those of TDDFT and BSE.
The active-space ppRPA approach has recently been developed to reduce the computational cost\cite{zhangAccurateEfficientCalculation2016}.
In the active-space ppRPA approach,
the ppRPA matrix is constructed and diagonalized in an active space,
which consists of a predetermined number of frontier orbitals near the 
highest occupied molecular orbital (HOMO) and the 
lowest unoccupied molecular orbital(LUMO)\cite{zhangAccurateEfficientCalculation2016}.
As shown in Ref.\citenum{zhangAccurateEfficientCalculation2016},
the convergence of excitation energies obtained from the active-space ppRPA approach toward the full-space ppRPA results is rapid with respect to the size of the active space.
The active-space ppRPA approach is shown to efficiently compute the excitation energies for different systems with only a modest number of active orbitals.
However,
a large amount of high-lying virtual orbitals are included in the original active-space ppRPA approach.
These high-lying virtual orbitals have small contributions to the desired low-lying excitations,
which lead to unnecessary computational burden and limit the application for larger systems.

In this work,
we introduce a new active-space ppRPA approach that improves the efficiency and further reduces the computational cost.
The new active-space ppRPA approach constrains both indexes in particle and hole pairs,
which selects frontier orbitals by the summation of orbital energies and is shown to provide significantly better efficiency than the original active-space ppRPA approach\cite{zhangAccurateEfficientCalculation2016}.
We also extend the active-space ppRPA approach to the T-matrix approximation\cite{zhangAccurateQuasiparticleSpectra2017,liRenormalizedSinglesGreen2021} for core-level QP energy calculations.
In the active-space T-matrix approach,
the self-energy in the active space is formulated with eigenvalues and eigenvectors obtained from the preceding active-space ppRPA approach.
To capture the contribution to the self-energy outside the active space,
we develop the non-interacting pair (NIP) approximation that only uses KS orbitals and orbital energies to approximate the ppRPA eigenvalues and eigenvectors.
We show that the active-space T-matrix approach combined with the recently developed RS Green's function\cite{jinRenormalizedSinglesGreen2019,liRenormalizedSinglesGreen2021,liBenchmarkGWMethods2022,liRenormalizedSinglesCorrelation2022},
which is shown to be a good starting point for the Green's function formalism,
is capable of predicting accurate core-level binding energies (CLBEs).

\section{THEORY}

\subsection{Excitation energies from the ppRPA}
We first review the ppRPA formalism.
Similar to the phRPA that is formulated in terms of the density fluctuation,
the ppRPA is formulated with the fluctuation of the pairing matrix\cite{vanaggelenExchangecorrelationEnergyPairing2013,vanaggelenExchangecorrelationEnergyPairing2014,ringNuclearManybodyProblem1980,ripkaQuantumTheoryFinite1986}
\begin{equation}
    \kappa (x_1, x_2) = \langle \Psi^N_0 | \hat{\psi} (x_2) \hat{\psi} (x_1) | \Psi^N_0 \rangle
\end{equation}
where $x=(r,\sigma)$ is the space-spin combined variable, $\Psi^N_0$ is the $N$-electron ground state, $\hat{\psi}^{\dagger}$ and $\hat{\psi}$ are the second quantization creation and annihilation operator.
While the paring matrix is zero for electronic ground states, 
the response of the pairing matrix $\delta \kappa (x_1, x_2)$ is non-zero when the system is perturbed by an external pairing field.
In the frequency space,
the time-ordered pairing matrix fluctuation, the linear response function, has poles at two-electron addition and removal energies\cite{vanaggelenExchangecorrelationEnergyPairing2013,vanaggelenExchangecorrelationEnergyPairing2014,ringNuclearManybodyProblem1980,ripkaQuantumTheoryFinite1986}
\begin{equation}
    K(\omega)_{pqrs} = \sum_m \frac{\langle \Psi^N_0 | \hat{a}_p \hat{a}_q | \Psi^{N+2}_0 \rangle \langle \Psi^{N+2}_0 | \hat{a}_s^{\dagger} \hat{a}_r^{\dagger} | \Psi^N_0 \rangle }{\omega - \Omega^{N+2}_m + i\eta }
    - \sum_m \frac{\langle \Psi^N_0 | \hat{a}_s^{\dagger} \hat{a}_r^{\dagger} | \Psi^{N-2}_0 \rangle \langle \Psi^{N-2}_0 | \hat{a}_p \hat{a}_q | \Psi^N_0 \rangle }{\omega - \Omega^{N-2}_m - i\eta }
\end{equation}
where  $\hat{a}_p^{\dagger}$ and $\hat{a}_p$ are the second quantization creation and annihilation operator for the orbital $p$, $\Omega^{N\pm2}$ is the two-electron addition/removal energy, $\eta$ is a positive infinitesimal number.
We use $i$, $j$, $k$, $l$ for occupied orbitals, $a$, $b$, $c$, $d$ for occupied orbitals, $p$, $q$, $r$, $s$ for general orbitals, and $m$ for the index of the two-electron addition/removal energy.
To obtain the pairing matrix fluctuation $K$ of the interacting system,
the ppRPA approximates $K$ in terms of the non-interacting $K_0$ by the Dyson equation\cite{vanaggelenExchangecorrelationEnergyPairing2013,vanaggelenExchangecorrelationEnergyPairing2014}
\begin{equation}\label{eq:dyson}
    K = K^0 + K^0 V K
\end{equation}
where the antisymmetrized interaction $V_{pqrs}=\langle pq || rs \rangle = \langle pq | rs \rangle - \langle pq | sr \rangle$ defined as 
\begin{align}
  \langle pq||rs\rangle & = \langle pq|rs\rangle-\langle pq|sr\rangle\\
 & =\int dx_{1}dx_{2}\frac{\phi_{p}^{*}(x_{1})\phi_{q}^{*}(x_{2})(1-\hat{P}_{12})\phi_{r}(x_{1})\phi_{s}(x_{2})}{|r_{1}-r_{2}|}.
\end{align}
The direct ppRPA can be obtained if the exchange term is neglected in $V$\cite{tahirComparingParticleparticleParticlehole2019}. 

Eq.~\ref{eq:dyson} can be cast into a generalized eigenvalue equation,
which describes the two-electron addition/removal excitations\cite{vanaggelenExchangecorrelationEnergyPairing2013,yangExcitationEnergiesParticleparticle2014,yangDoubleRydbergCharge2013}
\begin{equation}\label{eq:eigen_equation}
\begin{bmatrix}\mathbf{A} & \mathbf{B}\\
\mathbf{B}^{\text{T}} & \mathbf{C}
\end{bmatrix}\begin{bmatrix}\mathbf{X}\\
\mathbf{Y}
\end{bmatrix}=\Omega^{N\pm2}\begin{bmatrix}\mathbf{I} & \mathbf{0}\\
\mathbf{0} & \mathbf{-I}
\end{bmatrix}\begin{bmatrix}\mathbf{X}\\
\mathbf{Y}
\end{bmatrix}
\end{equation}
with
\begin{align}
A_{ab,cd} & =\delta_{ac}\delta_{bd}(\epsilon_{a}+\epsilon_{b})+\langle ab||cd\rangle \text{,}\\
B_{ab,kl} & =\langle ab||kl\rangle \text{,}\\
C_{ij,kl} & =-\delta_{ik}\delta_{jl}(\epsilon_{i}+\epsilon_{j})+\langle ij||kl\rangle \text{,}
\end{align}
where $a<b$, $c<d$, $i<j$, $k<l$ and $\Omega^{N\pm2}$ is the two-electron addition/removal energy.
To obtained the charged-neutral excitations of the $N$-electron system,
first the self-consistent field (SCF) calculation of the corresponding $(N-2)$-electron system at the same geometry is performed,
then the orbital energies and the orbitals are used in the working equation of the ppRPA Eq.~\ref{eq:eigen_equation} for two-electron addition energies.
The excitation energy can be obtained from the difference between the lowest and a higher two-electron addition energy\cite{yangBenchmarkTestsSpin2013}. 

\subsection{Active-space ppRPA approach for optical excitation energies}
To reduce the computational cost,
the active-space ppRPA approach has recently been developed.
The motivation is to approximate the resonance modes by choosing a set of frontier orbitals in an active space\cite{zhangAccurateEfficientCalculation2016}
\begin{equation}\label{eq:old_active}
    \delta \kappa (x_1, x_2, \Omega_m^{N\pm2} ) \approx
    \sum_{A<b} X^{N\pm2,m}_{Ab} \psi_A(x_1) \psi_b(x_2)
    + \sum_{i<J} Y^{N\pm2,m}_{iJ} \psi_i(x_1) \psi_J(x_2),
\end{equation}
where we use $A$, $B$ for active virtual orbitals and $I$, $J$ for active occupied orbitals.
In Eq.~\ref{eq:old_active},
only one index in particle and the hole pairs is constrained.
So the dimension of the active space introduced in Ref.\citenum{zhangAccurateEfficientCalculation2016} is
\begin{equation}\label{eq:dim_old_space}
    \frac{1}{2} (N_{\text{vir,act}} + 1) N_{\text{vir,act}} + N_{\text{vir,act}} (N_{\text{vir}} - N_{\text{vir,act}})
    + \frac{1}{2} (N_{\text{occ,act}} + 1) N_{\text{occ,act}} + N_{\text{occ,act}} (N_{\text{occ}} - N_{\text{occ,act}}),
\end{equation}
which depends on the size of the active space and linearly on the size of the full system.
As shown in Ref.\citenum{zhangAccurateEfficientCalculation2016},
this active-space ppRPA approach has small errors of $0.05$ \,{eV} for excitation energies toward the full-space ppRPA results with the scaling of $\mathcal{O}(N^4)$\cite{zhangAccurateEfficientCalculation2016}.
However,
this active-space ppRPA approach can still have challenges for calculating large systems,
because it only constrains one index in particle and hole pairs and includes high-lying virtual orbitals that have small contributions to the low-lying excitations. 

In this work,
we introduce a new active-space ppRPA approach to further reduce the computational cost.
The new active-space ppRPA approach constrains \textit{both} indexes in particle pairs and hole pairs,
which means
\begin{equation}
    \delta \kappa (x_1, x_2, \Omega_m^{N\pm2} ) \approx
    \sum_{A<B< N_{\text{vir,act}} } X^{N\pm2,m}_{AB} \psi_A(x_1) \psi_B(x_2)
    + \sum_{I<J< N_{\text{occ,act}} } Y^{N\pm2,m}_{IJ} \psi_I(x_1) \psi_J(x_2).
\end{equation}
Therefore, only low-lying virtual orbitals are included.
The dimension of the new active space is
\begin{equation}\label{eq:dim_new_space}
    \frac{1}{2} (N_{\text{vir,act}} + 1) N_{\text{vir,act}}
    + \frac{1}{2} (N_{\text{occ,act}} + 1) N_{\text{occ,act}},
\end{equation}
which only depends on the size of the active space, not on the system size.
Therefore, the scaling for the active-space ppRPA approach is $\mathcal{O}(N)$ when a constant number of orbitals are included in the active space for the construction of the three-center resolution of identity (RI) or density-fitting matrix is in the ground-state calculation.

\subsection{Active-space T-matrix approach for core-level quasiparticle energies}
The active-space ppRPA approach can be directly applied in the T-matrix approximation.
The T-matrix self-energy is the counterpart of the $GW$ self-energy in the particle-particle channel,
which is the summation of all ladder diagrams and is formulated with the ppRPA\cite{zhangAccurateQuasiparticleSpectra2017,liRenormalizedSinglesGreen2021}.
In the frequency space,
the correlation part of the self-energy in the T-matrix approximation is\cite{zhangAccurateQuasiparticleSpectra2017,liRenormalizedSinglesGreen2021}
\begin{equation}\label{eq:self_energy}
        \Sigma^{\text{c}}_{pq}(\omega) =
        \sum_m \sum_i \frac{\langle pi | \chi^{N+2}_m\rangle \langle qi | \chi^{N+2}_m\rangle}{\omega + \epsilon_i - \Omega^{N+2}_m + i\eta}
        + \sum_m \sum_a \frac{\langle pa | \chi^{N-2}_m\rangle \langle qa | \chi^{N-2}_m\rangle}{\omega + \epsilon_a - \Omega^{N-2}_m - i\eta}.
\end{equation}
In Eq.~\ref{eq:self_energy}, the transition density is
\begin{align}
    \langle pi | \chi^{N+2}_m\rangle =& \sum_{c<d}\langle pi||cd \rangle X^{N+2,m}_{cd} + \sum_{k<l}\langle pi||kl \rangle Y^{N+2,m}_{kl} \\
    \langle pa | \chi^{N-2}_m\rangle =& \sum_{c<d}\langle pa||cd \rangle X^{N-2,m}_{cd} + \sum_{k<l}\langle pa||kl \rangle Y^{N-2,m}_{kl},
\end{align}
where $X^{N\pm 2}_m$, $Y^{N\pm 2}_m$ and $\Omega^{N\pm 2}_m$ are two-electron addition/removal eigenvectors and eigenvalues from the ppRPA.
Since all excitations in the ppRPA are needed to formulate the T-matrix self-energy,
the scaling for the ppRPA step is of $\mathcal{O} (N^6)$.
The scaling of evaluating the T-matrix self-energy in Eq.~\ref{eq:self_energy} is also of $\mathcal{O} (N^6)$\cite{zhangAccurateQuasiparticleSpectra2017}. 

In our active-space T-matrix approach, the self-energy is divided into two parts:
the contribution from the active space and the contribution outside the active space.
For the contribution from the active space,
the correlation part of the self-energy is formulated with the eigenvalues and eigenvectors obtained from the active-space ppRPA approach
\begin{equation}
        \Sigma^{\text{c,act}}_{pq}(\omega) =
        \sum_m^{\text{act}} \sum_i^{ N_{\text{occ,act}} } \frac{\langle pi | \chi^{N+2}_m\rangle \langle qi | \chi^{N+2}_m\rangle}{\omega + \epsilon_i - \Omega^{N+2}_m - i\eta}
        + \sum_m^{\text{act}} \sum_a^{ N_{\text{vir,act}} } \frac{\langle pa | \chi^{N-2}_m\rangle \langle qa |     \chi^{N-2}_m\rangle}{\omega + \epsilon_a - \Omega^{N-2}_m + i\eta},
\end{equation}
where ``act" means the contribution from the active space.
The scaling of solving the ppRPA equation and evaluating the self-energy within the active space is of $\mathcal{O} (N^6)$, which is the same as the full-space T-matrix approach. 
However, because the indexes are constrained in the active space, a small prefactor can be obtained. \\

To include the contribution outside the active space, 
we introduce the NIP approximation.
In the NIP approximation,
for the excitation $m$ corresponding to the $(p,q)$ pair,
the two-electron addition/removal energy is simply approximated by the KS orbital energies
\begin{equation}
    \Omega_{pq}^{N\pm 2} = \pm (\epsilon_p + \epsilon_q),
\end{equation}
and the corresponding eigenvector is
\begin{equation}
    \chi_{pq}^{N\pm 2} = 1.
\end{equation}

Therefore, the correlation part of the self-energy from the NIP approximation is
\begin{equation}\label{eq:self_energy_nip}
        \Sigma^{\text{c,out}}_{pq}(\omega) =
        \sum_{jk}^{\text{out}} \sum_i \frac{\langle pi || jk \rangle \langle qi || jk \rangle}{\omega + \epsilon_i - (\epsilon_j + \epsilon_k) - i\eta}
        + \sum_{bc}^{\text{out}} \sum_a \frac{\langle pa || bc \rangle \langle qa || bc \rangle}{\omega + \epsilon_a - (\epsilon_b + \epsilon_c) + i\eta},
\end{equation}
where ``out" in the summations means indexes outside the active space.
The scaling of evaluating the self-energy with the NIP approximation in Eq.~\ref{eq:self_energy_nip} is of $\mathcal{O} (N^5)$.\\

\section{COMPUTATIONAL DETAILS}
We implemented the active-space ppRPA approach and the active-space T-matrix approach in the QM4D quantum chemistry package\cite{qm4d}.
For the active-space ppRPA approach we tested five different types of excitations:
S-T gaps of diradical systems\cite{yangSingletTripletEnergy2015},
CT excitation energies of the Stein CT test set\cite{steinReliablePredictionCharge2009},
double excitation energies of small atomic and molecular systems\cite{yangDoubleRydbergCharge2013},
Rydberg excitation energies of atomic systems\cite{xuTestingNoncollinearSpinFlip2014},
valence excitation energies of the Thiel test set\cite{silva-juniorBenchmarksElectronicallyExcited2008,schreiberBenchmarksElectronicallyExcited2008} and the Tozer test set\cite{peachExcitationEnergiesDensity2008}.
For S-T gaps of diradical systems,
the aug-cc-pVDZ basis set\cite{dunningGaussianBasisSets1989,kendallElectronAffinitiesFirst1992} was used.
Geometries and reference values were taken from Ref.\citenum{yangSingletTripletEnergy2015}.
For the adiabatic S-T gaps, 
corrections obtained from the SCF calculations of the $N$-electron system are added to vertical S-T gaps\cite{yangSingletTripletEnergy2015}
\begin{equation}\label{eq:adiabatic_st}
    E_{\text{g}_{\text{a}}} = E_{\text{g}_{\text{v,Sgeo}}} + (E_{\text{T,Sgeo}} - E_{\text{T,Tgeo}}),
\end{equation}
where $E_{\text{g}_{\text{a}}}$ is the adiabatic S-T gap, 
$E_{\text{g}_{\text{v,Sgeo}}}$ is the vertical S-T gap, 
$E_{\text{T,Sgeo}}$ is the triplet energy of the $N$-electron system of the singlet geometry and $E_{\text{T,Tgeo}}$ is the triplet energy of the $N$-electron system of the triplet geometry.
For the Stein CT test set\cite{steinReliablePredictionCharge2009},
the cc-pVDZ basis set\cite{dunningGaussianBasisSets1989} was used.
Geometries and experimental values in the gas phase were taken from Ref.\citenum{steinReliablePredictionCharge2009}.
For double excitations,
the even-tempered basis set defined in Ref.\citenum{yangDoubleRydbergCharge2013} was used for Be and Li,
the cc-pVQZ basis set\cite{dunningGaussianBasisSets1989} was used for BH,
the aug-cc-pVDZ basis set\cite{dunningGaussianBasisSets1989,kendallElectronAffinitiesFirst1992} was used for C and the cc-pVDZ basis set\cite{dunningGaussianBasisSets1989} was used for H in polyenes.
Geometries were optimized at 6-31G*/MP2 level with the GAUSSIAN16 A.03 software\cite{g16}.
Reference values were taken from Ref.\citenum{yangDoubleRydbergCharge2013}.
For Rydberg excitations,
the aug-cc-pVQZ basis set was used\cite{dunningGaussianBasisSets1989,kendallElectronAffinitiesFirst1992}.
Reference values were taken from Ref.\citenum{xuTestingNoncollinearSpinFlip2014}.
For the Thiel test set\cite{silva-juniorBenchmarksElectronicallyExcited2008,schreiberBenchmarksElectronicallyExcited2008} and the Tozer test set\cite{peachExcitationEnergiesDensity2008},
the aug-cc-pVDZ basis set\cite{dunningGaussianBasisSets1989,kendallElectronAffinitiesFirst1992} was used.
Geometries and reference values were taken from Ref.\citenum{silva-juniorBenchmarksElectronicallyExcited2008,schreiberBenchmarksElectronicallyExcited2008,peachExcitationEnergiesDensity2008}.
For the hydrocarbon diradical calculations,
the cc-pVDZ basis set\cite{dunningGaussianBasisSets1989} was used.
The closed-shell singlet geometries were taken from Ref.\citenum{escayolaEffectExocyclicSubstituents2019}.
Reference values were taken from Ref.\citenum{montgomeryMolecularStructuresThiele1986} and Ref.\citenum{schmidtEnergeticPositionsLowest1971}.
For the active-space T-matrix approach,
the CLBEs in the CORE65 test set\cite{golzeAccurateAbsoluteRelative2020} were calculated.
The def2-TZVP basis set\cite{weigendBalancedBasisSets2005} was used.
Geometries and reference values were taken from Ref.\citenum{golzeAccurateAbsoluteRelative2020}.
QM4D uses Cartesian basis sets and uses the RI technique\cite{weigendAccurateCoulombfittingBasis2006,renResolutionofidentityApproachHartree2012,eichkornAuxiliaryBasisSets1995} to compute two-electron integrals.
All basis sets and corresponding fitting basis sets\cite{weigendEfficientUseCorrelation2002,weigendRIMP2OptimizedAuxiliary1998} were taken from the Basis Set Exchange\cite{fellerRoleDatabasesSupport1996,schuchardtBasisSetExchange2007,pritchardNewBasisSet2019}.

\section{RESULTS}

\subsection{Comparison between different active-space ppRPA approaches}
\FloatBarrier

\begin{figure}
\includegraphics[width=0.9\textwidth]{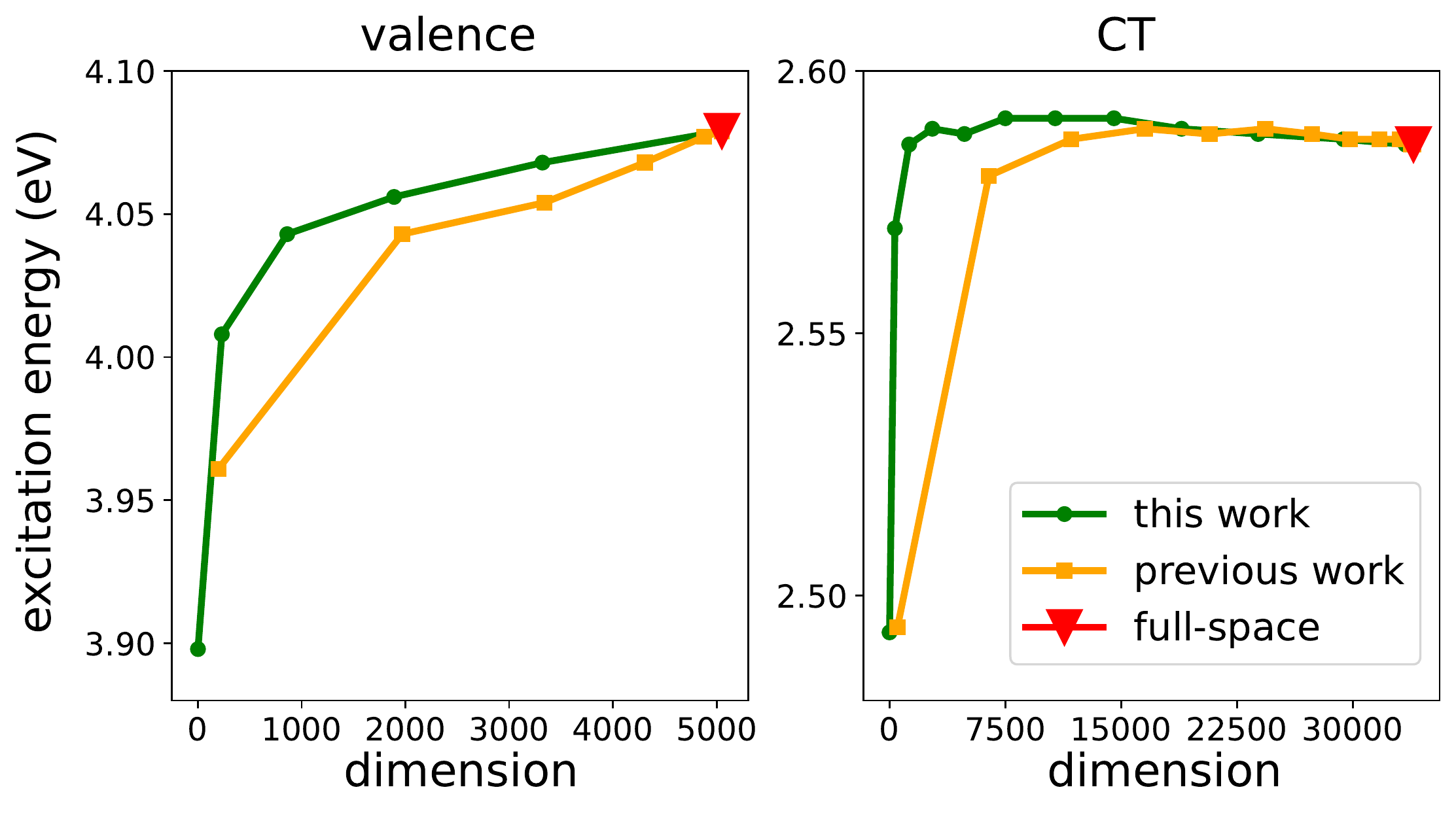}
\caption{Comparisons of excitation energies obtained from the active-space ppRPA approach in this work and in Ref.\citenum{zhangAccurateEfficientCalculation2016} with respect to the dimension of the ppRPA matrix.
The active spaces include all occupied orbitals and different numbers of virtual orbitals.
The dimensions of active spaces are calculated from Eq.~\ref{eq:dim_old_space} and Eq.~\ref{eq:dim_new_space}.
Left: the $^3$B$_2$ state of cyclopropene.
The full-space dimension is 5050.
Right: first singlet state of o-xylene.
The full-space dimension is 33930.
The aug-cc-pVDZ basis set was used for cyclopropene,
the cc-pVDZ basis set was used for o-xylene.}
\label{fig:pprpa_compare}
\end{figure}

We first compare the active-space ppRPA approach developed in this work and in Ref.\citenum{zhangAccurateEfficientCalculation2016}.
The excitation energies of the $^3$B$_2$ of cyclopropene and first singlet state of o-xylene obtained by different active-space ppRPA approaches are shown in Fig.~\ref{fig:pprpa_compare}.
For simplicity,
the active spaces include all occupied orbitals and the varying number of virtual orbitals.
As shown in the left of Fig.~\ref{fig:pprpa_compare},
the active-space ppRPA approach in this work has better convergence behavior than that in Ref.\citenum{zhangAccurateEfficientCalculation2016}.
The reason is the active-space ppRPA approach in Ref.\citenum{zhangAccurateEfficientCalculation2016} constrains only one index in particle pairs,
which includes high-lying particle pairs with small contributions to the low-lying excitation energies.
As shown on the right-hand side of Fig.~\ref{fig:pprpa_compare},
for the CT excitation energy,
both active-space ppRPA approaches with a small number of virtual orbitals in the active space produce errors smaller than $0.05$ \,{eV} compared with the full-space results.
However,
as shown in Eq.~\ref{eq:dim_old_space},
the size of the active space in Ref.\citenum{zhangAccurateEfficientCalculation2016} depends on the size of the full system.
Therefore,
the dimension can be large even with a small number of active orbitals.
For the active space in this work, 
only particle and hole pairs with dominant contributions to the low-lying excitation energies are included,
which lead to a small dimension.
Thus,
the active-space ppRPA approach in this work is more efficient to predict excitation energies.

\subsection{Excitation energies from active-space ppRPA}

\subsubsection{S-T gaps of diradicals}

\begin{table}
\setlength\tabcolsep{2pt}
\caption{\label{tab:diradical}Adiabatic singlet-triplet gaps and MAEs of diradicals obtained from the full-space and the active-space ppRPA approaches based on HF, PBE and B3LYP. 
full stands for the full-space.
(20,20)/(30,30) means 20/30 occupied and virtual frontier orbitals are included in the active space.
Geometries and reference values were taken from Ref.\citenum{yangSingletTripletEnergy2015}.
The aug-cc-pVDZ basis set was used.
All values in kcal/mol.}
\begin{tabular}{cccccccccccc}
\toprule
               &            &       & \multicolumn{3}{c}{HF}    & \multicolumn{3}{c}{PBE}   & \multicolumn{3}{c}{B3LYP} \\
\cmidrule(l{0.5em}r{0.5em}){4-6} \cmidrule(l{0.5em}r{0.5em}){7-9} \cmidrule(l{0.5em}r{0.5em}){10-12}
               & state      & Ref   & full  & (30,30) & (20,20) & full  & (30,30) & (20,20) & full  & (30,30) & (20,20) \\
\midrule
$\cdot$\ce{CH2CH2CH2}$\cdot$ &            & 1.8   & 2.4   & 2.7     & 3.1     & 5.4   & 5.9     & 5.9     & 4.4   & 4.8     & 4.8     \\
$\cdot$\ce{CH2(CH2)4CH2}$\cdot$           & $^1$A$_1$  & 0.0   & 0.0   & 0.0     & 0.0     & 0.1   & 0.1     & 0.1     & 0.1   & 0.1     & 0.1     \\
               & $^1$B$_1$  & 160.8 & 79.9  & 79.8    & 79.8    & 159.2 & 159.4   & 159.5   & 147.4 & 147.5   & 147.6   \\
               & 2$^1$A$_1$ & 163.1 & 80.6  & 80.5    & 80.5    & 162.0 & 162.0   & 162.5   & 149.0 & 149.1   & 149.5   \\
$\cdot$\ce{CH2(CH2)4C(CH3)H}$\cdot$           & $^1$A$_1$  & -0.2  & -0.3  & -0.3    & -0.3    & 0.4   & 0.4     & 0.4     & 0.1   & 0.1     & 0.1     \\
               & 2$^1$A$_1$ & 131.1   & 66.1    & 65.6    & 65.6    & 139.8 & 139.7   & 140.2   & 129.1 & 129.0   & 129.2   \\
               & 3$^1$A$_1$ & 144.3   & 78.3    & 77.7    & 78.0    & 152.4 & 152.4   & 152.5   & 142.1 & 142.1   & 142.0   \\
cyclobutadiene &            & 7.2     & 7.0     & 7.8     & 6.5     & 6.4   & 6.9     & 5.3     & 6.5   & 7.1     & 5.6     \\
NH             &            & 35.9    & 30.9    & 30.9    & 31.0    & 40.5  & 40.5    & 40.8    & 38.5  & 38.5    & 38.7    \\
\ce{OH+}       &            & 50.5    & 45.5    & 45.5    & 45.7    & 54.2  & 54.2    & 54.5    & 52.2  & 52.3    & 52.6    \\
NF             &            & 34.3    & 28.5    & 29.4    & 29.0    & 28.3  & 29.1    & 28.7    & 28.6  & 29.5    & 29.1    \\
\ce{O2}        &            & 22.6    & 23.1    & 23.3    & 24.7    & 23.5  & 23.7    & 24.9    & 23.6  & 23.8    & 25.0    \\
\ce{CH2}       & $^1$A$_1$  & 9.7     & $-$1.9  & $-$1.3  & $-$0.7    & 6.3   & 7.2     & 8.0     & 4.2   & 5.1     & 5.8     \\
               & $^1$B$_1$  & 32.5    & 27.4    & 27.7    & 27.1    & 34.6  & 35.1    & 34.8    & 33.2  & 33.6    & 33.3    \\
               & 2$^1$A$_1$ & 58.3    & 49.3    & 49.8    & 48.8    & 58.8  & 59.6    & 59.3    & 56.9  & 57.7    & 57.3    \\
\ce{NH2+}      & $^1$A$_1$  & 28.9    & 12.8    & 14.7    & 15.3    & 25.6  & 27.8    & 28.2    & 23.0  & 25.2    & 25.6    \\
               & $^1$B$_1$  & 43.0    & 36.4    & 36.9    & 38.0    & 43.7  & 44.3    & 45.2    & 42.0  & 42.6    & 43.6    \\
               & 2$^1$A$_1$ & 76.5    & 62.1    & 65.5    & 66.8    & 72.3  & 76.4    & 77.9    & 70.0  & 74.0    & 75.5    \\
\ce{SiH2}      & $^1$A$_1$  & $-$20.6 & $-$25.0 & $-$24.5 & $-$25.0   & $-$28.4 & $-$27.7   & $-$26.9   & $-$28.8 & $-$28.2   & $-$27.3   \\
               & $^1$B$_1$  & 24.1    & 24.0    & 24.0    & 25.4    & 26.4  & 26.5    & 27.5    & 26.2  & 26.3    & 27.4    \\
               & 2$^1$A$_1$ & 57.0    & 51.3    & 51.5    & 54.3    & 52.2  & 52.5    & 54.6    & 53.5  & 53.8    & 55.9    \\
\ce{PH2+}      & $^1$A$_1$  & $-$18.3 & $-$25.3 & $-$24.7 & $-$23.8   & $-$23.6 & $-$23.0   & $-$22.3   & $-$24.7 & $-$24.0   & $-$23.3   \\
               & $^1$B$_1$  & 27.6    & 28.8    & 29.1    & 30.3    & 30.1  & 30.3    & 31.0    & 30.0  & 30.2    & 31.0    \\
               & 2$^1$A$_1$ & 65.6    & 60.1    & 60.7    & 62.2    & 59.4  & 59.7    & 62.5    & 60.6  & 60.9    & 63.9    \\
\midrule
MAE            &            &       & 36.9  & 37.2    & 37.2    & 3.1   & 3.0     & 3.2     & 4.4   & 4.4     & 4.5     \\
\bottomrule
\end{tabular}
\end{table}

We then examined the performance of the active-space ppRPA approach on calculating the S-T gaps of diradicals. 
Adiabatic S-T gaps and mean absolute errors (MAEs) of four diatomic diradicals and four carbene-like diradicals, vertical S-T gaps of three disjoint diradicals and one four-$\pi$-electron diradical obtained from the fulls-space and the active-space ppRPA approach based on HF, PBE and B3LYP are tabulated in Table.~\ref{tab:diradical}. 
The vertical S-T gaps were directly calculated as the difference between the two-electron addition energies of the singlet state and that of the triplet ground state.
The adiabatic S-T gaps were obtained using the correction defined in Eq.~\ref{eq:adiabatic_st}.
It can be seen that by using an active-space consisting of 30 occupied and 30 virtual orbitals, 
the active-space ppRPA approach based on both the HF and the DFT references provides errors smaller than $0.3$ kcal/mol compared with the full-space results,
which are smaller than the errors of $1.0$ kcal/mol by using the active space in Ref.\citenum{zhangAccurateEfficientCalculation2016}. 
For carbene-like diradicals \ce{NH2+}
the error of the S-T gap for the 2$^1$A$_1$ state is $4.0$ kcal/mol,
which means the size of the active space needs to be increased for S-T gaps of high-lying states.
As shown in Table.~\ref{tab:diradical},
using an active-space consisting of a constant number of active orbitals, 
all S-T gaps obtained from the ppRPA approach have errors within $0.5$ kcal/mol compared with the full-space results.

\subsubsection{CT excitation energy}

\begin{table}
\setlength\tabcolsep{3.5pt}
\caption{\label{tab:ct}CT excitation energies and MAEs in the Stein CT test set obtained from the full-space and the active-space ppRPA approaches based on HF, PBE and B3LYP. 
full stands for the full-space.
(20,20)/(30,30) means 20/30 occupied and virtual frontier orbitals are included in the active space.
Geometries were taken from Ref.\citenum{steinReliablePredictionCharge2009}.
Experimental values in the gas phase were taken as the reference values\cite{steinReliablePredictionCharge2009}.
The cc-pVDZ basis set was used.
All values in eV.}
\begin{tabular}{ccccccccccc}
\toprule
                    &        & \multicolumn{3}{c}{HF}   & \multicolumn{3}{c}{PBE}  & \multicolumn{3}{c}{B3LYP} \\
\cmidrule(l{0.5em}r{0.5em}){3-5} \cmidrule(l{0.5em}r{0.5em}){6-8} \cmidrule(l{0.5em}r{0.5em}){9-11}
                    & Ref    & full & (30,30) & (20,20) & full & (30,30) & (20,20) & full  & (30,30) & (20,20) \\
\midrule
anthracene          & 2.05   & 1.95 & 1.92    & 1.91    & 1.25 & 1.21    & 1.20    & 1.36  & 1.32    & 1.32    \\
9-cyano             & 2.33   & 1.99 & 1.93    & 1.92    & 1.35 & 1.34    & 1.31    & 1.49  & 1.45    & 1.42    \\
9-cholo             & 2.06   & 1.91 & 1.87    & 1.86    & 1.19 & 1.16    & 1.13    & 1.31  & 1.26    & 1.23    \\
9-carbo-methoxy     & 2.16   & 1.92 & 1.87    & 1.86    & 1.18 & 1.13    & 1.10    & 1.30  & 1.25    & 1.22    \\
9-methyl            & 1.87   & 1.70 & 1.67    & 1.66    & 1.05 & 1.02    & 0.99    & 1.15  & 1.12    & 1.10    \\
9,10-dimethyl       & 1.76   & 1.78 & 1.77    & 1.76    & 1.07 & 1.05    & 1.04    & 1.19  & 1.18    & 1.16    \\
9-formyl            & 2.22   & 2.08 & 2.04    & 2.03    & 1.43 & 1.40    & 1.37    & 1.50  & 1.46    & 1.44    \\
9-formyl, 10-chloro & 2.28   & 2.13 & 2.09    & 2.07    & 1.41 & 1.38    & 1.35    & 1.51  & 1.47    & 1.44    \\
benzene             & 3.91   & 2.58 & 2.38    & 2.36    & 3.18 & 3.23    & 3.23    & 3.46  & 3.50    & 3.55    \\
toluene             & 3.68   & 2.30 & 2.17    & 2.15    & 2.77 & 2.81    & 2.80    & 2.99  & 3.00    & 3.03    \\
o-xylene            & 3.47   & 1.84 & 1.68    & 1.63    & 2.41 & 2.43    & 2.42    & 2.59  & 2.57    & 2.58    \\
naphthalene         & 2.92   & 2.24 & 2.14    & 2.13    & 2.02 & 2.02    & 2.04    & 2.17  & 2.14    & 2.14    \\
\midrule
MAE                 &        & 0.53 & 0.60    & 0.61    & 0.87 & 0.88    & 0.89    & 0.72  & 0.75    & 0.76    \\
\bottomrule
\end{tabular}
\end{table}

We further investigated the performance of the active-space ppRPA approach for predicting CT excitation energies.
The first singlet excitation energies and MAEs of 12 intramolecular CT systems in the Stein CT test set obtained from the fulls-space and the active-space ppRPA approach based on HF, PBE and B3LYP are tabulated in Table.~\ref{tab:ct}.
It shows that for the active-space ppRPA approach with an active space consisting of 30 occupied orbitals and 30 virtual orbitals provides errors smaller than $0.1$ \,{eV} compared with the full-space ppRPA approach.
The active-space ppRPA approach based on both the generalized gradient approximation (GGA) functional and the hybrid functional B3LYP has better convergence performance than that based on the HF reference, 
which  has errors smaller than $0.03$ \,{eV}.
As shown in Table.~\ref{tab:ct},
using an active-space consisting of a constant number of active orbitals, 
all excitation energies of CT systems obtained from the active-space ppRPA@PBE and the active-space ppRPA@B3LYP approaches have errors within $0.05$ \,{eV} compared with the full-space results.

\subsubsection{Double excitations}

\begin{table}
\setlength\tabcolsep{3.5pt}
\caption{\label{tab:double}Double excitation energies and MAEs of small molecules obtained from the full-space and the active-space ppRPA approaches based on HF, PBE and B3LYP. 
full stands for the full-space.
(20,20)/(30,30) means 20/30 occupied and virtual frontier orbitals are included in the active space.
Geometries were optimized at 6-31G*/MP2 level with the GAUSSIAN16 A.03 software\cite{g16}.
Reference values were taken from Ref.\citenum{yangDoubleRydbergCharge2013}.
The even-tempered basis set defined in Ref.\citenum{yangDoubleRydbergCharge2013} was used for Be and Li,
the cc-pVQZ basis set was used for BH,
the aug-cc-pVDZ basis set was used for C and cc-pVDZ basis set was used for H in polyenes.
All values in eV.}
\begin{tabular}{cccccccccccc}
\toprule
           &                      &      & \multicolumn{3}{c}{HF}   & \multicolumn{3}{c}{PBE}  & \multicolumn{3}{c}{B3LYP} \\
\cmidrule(l{0.5em}r{0.5em}){4-6} \cmidrule(l{0.5em}r{0.5em}){7-9} \cmidrule(l{0.5em}r{0.5em}){10-12}
           & state                & Ref  & full & (30,30) & (20,20) & full & (30,30) & (20,20) & full  & (30,30) & (20,20) \\
\midrule
Be         & $^{1}$D              & 7.05 & 7.06 & 7.03    & 7.03    & 7.61 & 7.69    & 7.70    & 7.96  & 8.05    & 8.07    \\
           & $^{3}$P              & 7.40 & 7.45 & 7.46    & 7.47    & 7.49 & 7.52    & 7.54    & 7.84  & 7.88    & 7.90    \\
BH         & $^{3}$$\Sigma$       & 5.04 & 5.52 & 5.46    & 5.41    & 4.89 & 4.84    & 4.80    & 5.12  & 5.07    & 5.02    \\
           & $^{1}$$\Delta$       & 6.06 & 6.15 & 6.12    & 6.12    & 5.80 & 5.80    & 5.83    & 5.97  & 5.96    & 5.99    \\
           & $^{1}$$\Sigma$       & 7.20 & 7.08 & 7.06    & 7.10    & 6.90 & 6.94    & 7.04    & 7.04  & 7.07    & 7.16    \\
Butadiene  & $^{1}$A$_{\text{g}}$ & 6.55 & 5.92 & 5.82    & 5.88    & 6.12 & 6.10    & 6.14    & 6.45  & 6.43    & 6.48    \\
Hexatriene & $^{1}$A$_{\text{g}}$ & 5.21 & 5.43 & 5.38    & 5.39    & 4.49 & 4.48    & 4.48    & 5.00  & 4.99    & 5.01    \\
\midrule
MAE        &                      &      & 0.23 & 0.23    & 0.21    & 0.36 & 0.38    & 0.37    & 0.28  & 0.30    & 0.28     \\
\bottomrule
\end{tabular}
\end{table}

We then move on to the performance of the active-space ppRPA approach for predicting double excitation energies.
The ppRPA approach can naturally describe the double excitation energies.
The double excitation energies and MAEs of four molecules obtained from the fulls-space and the active-space ppRPA approach based on HF, PBE and B3LYP are tabulated in Table.~\ref{tab:double}.
It shows that with an active space consisting of 30 occupied and 30 virtual orbitals,
the active-space ppRPA approach based on all references provides errors smaller than $0.02$ \,{eV} compared with the full space results, 
which are smaller than the errors of $0.05$ \,{eV} in Ref.\citenum{zhangAccurateEfficientCalculation2016}.
As shown in Table.~\ref{tab:double},
an active space consisting a constant number of orbitals provides small errors for all tested systems.

\subsubsection{Rydberg excitations}

\begin{table}\setlength\tabcolsep{3pt}
\caption{\label{tab:rydberg}Rydberg excitation energies and MAEs of atomic systems obtained from the full-space and the active-space ppRPA approaches based on HF, PBE and B3LYP. 
full stands for the full-space.
(20,20)/(30,30) means 20/30 occupied and virtual frontier orbitals are included in the active space.
Reference values were taken from Ref.\citenum{xuTestingNoncollinearSpinFlip2014}.
The aug-cc-pVQZ basis set was used.
All values in eV.}
\begin{tabular}{cccccccccccc}
\toprule
    &                   &       & \multicolumn{3}{c}{HF}    & \multicolumn{3}{c}{PBE}   & \multicolumn{3}{c}{B3LYP} \\
\cmidrule(l{0.5em}r{0.5em}){4-6} \cmidrule(l{0.5em}r{0.5em}){7-9} \cmidrule(l{0.5em}r{0.5em}){10-12}
    & state             & Ref   & full  & (30,30) & (20,20) & full  & (30,30) & (20,20) & full  & (30,30) & (20,20) \\
\midrule
Be       & triplet 2s$\to$3s & 6.46  & 6.44  & 6.35    & 6.35    & 7.93  & 7.79    & 7.79    & 8.29  & 8.16    & 8.15    \\
         & singlet 2s$\to$3s & 6.78  & 6.77  & 6.70    & 6.70    & 8.22  & 8.12    & 8.12    & 8.59  & 8.51    & 8.50    \\
\ce{B+}  & triplet 2s$\to$3s & 16.09 & 16.06 & 15.95   & 15.96   & 18.60 & 18.45   & 18.46   & 18.51 & 18.37   & 18.39   \\
         & singlet 2s$\to$3s & 17.06 & 17.09 & 17.28   & 17.34   & 19.42 & 19.41   & 19.47   & 19.49 & 19.56   & 19.65   \\
Mg       & triplet 3s$\to$4s & 5.11  & 5.01  & 4.96    & 4.94    & 7.05  & 6.98    & 6.97    & 7.06  & 6.99    & 6.97    \\
         & singlet 3s$\to$4s & 5.39  & 5.29  & 5.26    & 5.25    & 7.32  & 7.27    & 7.26    & 7.31  & 7.27    & 7.26    \\
\ce{Al+} & triplet 3s$\to$4s & 11.32 & 11.14 & 11.09   & 11.36   & 14.09 & 14.03   & 14.31   & 14.29 & 14.33   & 14.52   \\
         & singlet 3s$\to$4s & 11.82 & 11.64 & 11.64   & 12.35   & 14.57 & 14.57   & 15.23   & 14.75 & 14.75   & 15.38   \\
\midrule
MAE      &                   &       & 0.08  & 0.15    & 0.18    & 2.15  & 2.07    & 2.20    & 2.28  & 2.24    & 2.35    \\
\bottomrule
\end{tabular}
\end{table}

We then evaluated the efficiency of the active-space ppRPA approach for predicting Rydberg excitation energies.
The Rydberg excitation energies and MAEs of four atoms obtained from the fulls-space and the active-space ppRPA approach based on HF, PBE and B3LYP are shown in Table.~\ref{tab:rydberg}. 
With an active space consisting of 30 occupied and 30 virtual orbitals,
the active-space ppRPA approach based on all references provides errors smaller than $0.1$ \,{eV} compared with the full space results, 
which is lower than the errors of $0.2$ \,{eV} in Ref.\citenum{zhangAccurateEfficientCalculation2016}.
The active-space ppRPA@HF approach gives the smallest MAE around $0.15$ \,{eV},
because HF provides the correct long-range behavior for describing Rydberg states.
Because Rydberg excitations are typically high-lying excited states, 
a larger active space is needed to obtain errors smaller than $0.05$ \,{eV}.
However, 
an active space containing a constant number of orbitals provides errors smaller than $0.1$ \,{eV} for all tested systems.

\subsubsection{Valence excitations}

\begin{table}
\setlength\tabcolsep{2.5pt}
\caption{\label{tab:valence}Valence excitation energies and MAEs of the Thiel test set\cite{silva-juniorBenchmarksElectronicallyExcited2008,schreiberBenchmarksElectronicallyExcited2008} and the Tozer test set\cite{peachExcitationEnergiesDensity2008} energies obtained from the full-space and the active-space ppRPA approaches based on HF, PBE and B3LYP. 
full stands for the full-space.
(20,20)/(30,30) means 20/30 occupied and virtual frontier orbitals are included in the active space.
Geometries and reference values were taken from Ref.\citenum{silva-juniorBenchmarksElectronicallyExcited2008,schreiberBenchmarksElectronicallyExcited2008,peachExcitationEnergiesDensity2008}.
The aug-cc-pVDZ basis set was used.
All values in eV.}
\begin{tabular}{cccccccccccc}
\toprule
                &              &      & \multicolumn{3}{c}{HF}   & \multicolumn{3}{c}{PBE}  & \multicolumn{3}{c}{B3LYP} \\
\cmidrule(l{0.5em}r{0.5em}){4-6} \cmidrule(l{0.5em}r{0.5em}){7-9} \cmidrule(l{0.5em}r{0.5em}){10-12}
                & state        & Ref  & full & (30,30) & (20,20) & full & (30,30) & (20,20) & full  & (30,30) & (20,20) \\
\midrule
ethene          & $^3$B$_{1u}$ & 4.50 & 3.92 & 3.81    & 3.76    & 3.48 & 3.40    & 3.40    & 3.62  & 3.53    & 3.52    \\
ethene          & $^1$B$_{1u}$ & 7.80 & 6.25 & 6.02    & 5.88    & 8.86 & 8.93    & 9.09    & 8.46  & 8.43    & 8.31    \\
butadiene       & $^3$B$_{u}$  & 3.20 & 3.21 & 3.13    & 3.13    & 2.28 & 2.20    & 2.23    & 2.52  & 2.44    & 2.46    \\
butadiene       & $^3$A$_{g}$  & 5.08 & 5.60 & 5.27    & 5.47    & 5.85 & 5.82    & 5.87    & 6.07  & 6.03    & 6.09    \\
butadiene       & $^1$B$_{u}$  & 6.18 & 5.49 & 4.55    & 4.56    & 6.52 & 6.76    & 6.92    & 6.58  & 6.77    & 6.85    \\
butadiene       & $^1$A$_{g}$  & 6.55 & 5.92 & 5.27    & 5.43    & 6.11 & 6.09    & 6.14    & 6.44  & 6.42    & 6.47    \\
hexatriene      & $^3$B$_{u}$  & 2.40 & 2.58 & 2.53    & 2.57    & 1.61 & 1.56    & 1.55    & 1.88  & 1.83    & 1.83    \\
hexatriene      & $^3$A$_{g}$  & 4.15 & 5.20 & 3.90    & 3.98    & 4.47 & 4.46    & 4.46    & 4.86  & 4.86    & 4.87    \\
hexatriene      & $^1$A$_{g}$  & 5.09 & 5.42 & 3.92    & 4.00    & 4.49 & 4.48    & 4.47    & 4.99  & 4.99    & 4.99    \\
hexatriene      & $^1$B$_{u}$  & 5.10 & 5.04 & 4.09    & 4.08    & 5.14 & 5.29    & 5.38    & 5.30  & 5.44    & 5.55    \\
octetraene      & $^3$B$_{u}$  & 2.20 & 2.15 & 2.10    & 2.10    & 1.21 & 1.18    & 1.16    & 1.49  & 1.46    & 1.44    \\
octetraene      & $^3$A$_{g}$  & 3.55 & 4.75 & 3.69    & 3.75    & 3.57 & 3.55    & 3.54    & 4.00  & 3.99    & 3.99    \\
octetraene      & $^1$A$_{g}$  & 4.47 & 5.02 & 3.70    & 3.77    & 3.53 & 3.52    & 3.51    & 4.08  & 4.07    & 4.07    \\
octetraene      & $^1$B$_{u}$  & 4.66 & 4.58 & 3.78    & 3.77    & 4.30 & 4.44    & 4.47    & 4.50  & 4.65    & 4.65    \\
cyclopropene    & $^3$B$_2$    & 4.34 & 4.22 & 4.08    & 4.05    & 3.91 & 3.86    & 3.85    & 4.08  & 4.03    & 4.01    \\
cyclopropene    & $^1$B$_2$    & 7.06 & 5.87 & 4.29    & 4.22    & 7.33 & 7.50    & 7.54    & 7.31  & 7.44    & 7.46    \\
cyclopentadiene & $^3$B$_2$    & 3.25 & 3.12 & 3.02    & 2.99    & 2.52 & 2.46    & 2.49    & 2.66  & 2.58    & 2.62    \\
cyclopentadiene & $^3$A$_1$    & 5.09 & 5.45 & 3.16    & 3.12    & 5.07 & 5.04    & 5.10    & 5.32  & 5.29    & 5.38    \\
cyclopentadiene & $^1$B$_2$    & 5.55 & 5.16 & 3.19    & 3.15    & 5.46 & 5.68    & 5.96    & 5.47  & 5.68    & 5.79    \\
cyclopentadiene & $^1$A$_1$    & 6.31 & 6.01 & 3.64    & 3.80    & 6.16 & 6.18    & 6.26    & 6.42  & 6.44    & 6.55    \\
norbornadiene   & $^3$A$_2$    & 3.72 & 3.77 & 3.39    & 3.50    & 3.58 & 3.57    & 3.54    & 3.68  & 3.66    & 3.64    \\
norbornadiene   & $^3$B$_2$    & 4.16 & 4.48 & 3.64    & 3.75    & 4.51 & 4.49    & 4.45    & 4.75  & 4.72    & 4.68    \\
norbornadiene   & $^1$A$_2$    & 5.34 & 3.95 & 3.41    & 3.54    & 5.35 & 5.52    & 5.72    & 5.36  & 5.54    & 5.71    \\
norbornadiene   & $^1$B$_2$    & 6.11 & 4.58 & 3.91    & 4.21    & 7.00 & 7.12    & 7.25    & 6.78  & 6.69    & 6.90    \\
furan           & $^3$B$_2$    & 4.17 & 3.82 & 3.65    & 3.64    & 3.37 & 3.29    & 3.30    & 3.49  & 3.41    & 3.41    \\
furan           & $^3$A$_1$    & 5.48 & 5.78 & 3.69    & 3.72    & 5.12 & 5.07    & 5.13    & 5.37  & 5.32    & 5.39    \\
furan           & $^1$B$_2$    & 6.32 & 5.09 & 3.68    & 3.67    & 6.58 & 6.69    & 6.82    & 6.50  & 6.38    & 6.42    \\
furan           & $^1$A$_1$    & 6.57 & 6.61 & 4.08    & 4.08    & 6.65 & 6.75    & 6.82    & 6.88  & 6.91    & 7.09    \\
s-tetrazine     & $^3$B$_{3u}$ & 1.89 & 3.28 & 3.24    & 3.28    & 1.86 & 1.82    & 1.85    & 2.19  & 2.14    & 2.18    \\
s-tetrazine     & $^3$A$_{u}$  & 3.52 & 5.38 & 5.32    & 5.38    & 3.16 & 3.12    & 3.09    & 3.67  & 3.62    & 3.61    \\
s-tetrazine     & $^1$B$_{3u}$ & 2.24 & 3.78 & 3.77    & 3.85    & 2.41 & 2.42    & 2.45    & 2.73  & 2.73    & 2.78    \\
s-tetrazine     & $^1$A$_{u}$  & 3.48 & 5.60 & 5.55    & 5.64    & 3.40 & 3.37    & 3.35    & 3.91  & 3.87    & 3.87    \\
formaldehyde    & $^3$A$_2$    & 3.50 & 1.66 & 1.50    & 1.39    & 3.41 & 3.32    & 3.26    & 3.15  & 3.05    & 2.98    \\
formaldehyde    & $^1$A$_2$    & 3.88 & 2.00 & 1.85    & 1.75    & 3.97 & 3.90    & 3.85    & 3.68  & 3.60    & 3.54    \\
acetone         & $^3$A$_2$    & 4.05 & 3.10 & 2.36    & 2.30    & 4.24 & 4.17    & 4.29    & 4.14  & 4.04    & 4.18    \\
acetone         & $^1$A$_2$    & 4.40 & 3.38 & 2.39    & 2.34    & 4.66 & 4.59    & 4.72    & 4.55  & 4.46    & 4.62    \\
benzoquinone    & $^3$B$_{1g}$ & 2.51 & 4.75 & 4.76    & 4.76    & 2.45 & 2.45    & 2.47    & 2.93  & 2.94    & 2.95    \\
benzoquinone    & $^1$B$_{1g}$ & 2.78 & 4.98 & 5.02    & 5.02    & 2.65 & 2.67    & 2.70    & 3.14  & 3.17    & 3.19    \\
\midrule
MAE             &              &      & 0.85 & 1.36    & 1.36    & 0.39 & 0.42    & 0.46    & 0.37  & 0.39    & 0.43    \\
\bottomrule
\end{tabular}
\end{table}

In this section, we investigate the effectiveness of the active-space ppRPA method in providing accurate valence excitation energies.
The valence excitation energies and MAEs of molecules in the Thiel test set and the Tozer test set obtained from the fulls-space and the active-space ppRPA approach based on HF, PBE and B3LYP are tabulated in Table.~\ref{tab:valence}.
With an active space consisting of 30 occupied and 30 virtual orbitals, 
the active-space ppRPA approach based on the HF reference has errors larger than $0.5$ \,{eV}.
The active-space ppRPA approach based on different DFT references has much smaller MAEs about $0.03$ \,{eV} compared with full-space results,
which are smaller than the errors of $0.1$ \,{eV} in Ref.\citenum{zhangAccurateEfficientCalculation2016}.
Results in Table.~\ref{tab:valence} show that
with DFT references, 
an active space consisting a constant number of orbitals provides errors around $0.05$ \,{eV} for all tested systems.

\subsubsection{S-T gaps of hydrocarbons}
\begin{figure}
\includegraphics[width=0.5\textwidth]{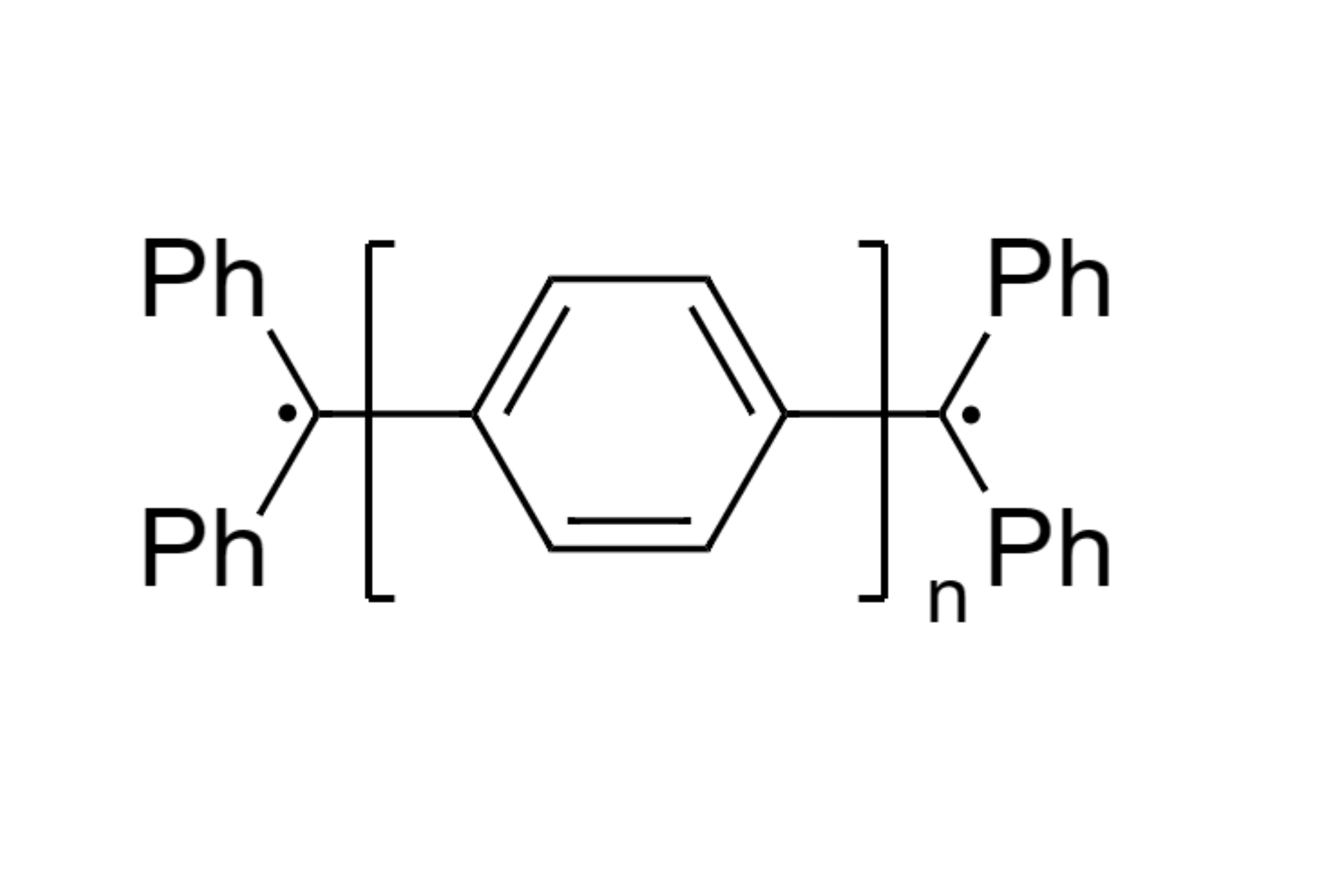}
\caption{Structures of hydrocarbons. 
$n=2$: the Chichibabin's hydrocarbon.
$n=3$: the M\"{u}ller's hydrocarbon.}
\label{fig:hydrocarbon}
\end{figure}

To demonstrate the performance of the active-space ppRPA approach,
we calculated the S-T gaps of the Chichibabin's hydrocarbon\cite{tschitschibabinUberEinigePhenylierte1907} and the M\"{u}ller's hydrocarbon\cite{mullerUberBiradikaloideTerphenylderivate1941} (as shown in Fig.~\ref{fig:hydrocarbon}), 
which contains 66 atoms and 76 atoms, 
respectively.
These two hydrocarbons show a strong reactivity and diradical character because of the formation of aromatic rings\cite{abeDiradicals2013}.
The M\"{u}ller's hydrocarbon is expected to have a stronger diradical character and a smaller S-T gap than those of the Chichibabin's hydrocarbon because it has one more aromatic ring\cite{abeDiradicals2013,schmidtEnergeticPositionsLowest1971,escayolaEffectExocyclicSubstituents2019}.
The S-T gaps and the dominant configuration contributions (DCC) of two hydrocarbons obtained from the active-space ppRPA@B3LYP approach with the cc-pVDZ basis set are shown in Table.~\ref{tab:hydrocarbon}.
30 occupied and 30 virtual orbitals are included in the active space.
The calculated S-T gap of the Chichibabin's hydrocarbon is $-3.0$ kcal/mol, 
which is close to the experimental value of $-5.5$ kcal/mol in Ref.\citenum{montgomeryMolecularStructuresThiele1986}.
The active-space ppRPA approach predicts a smaller S-T gap of $-0.1$ kcal/mol for the M\"{u}ller's hydrocarbon, 
which is similar to the reference value of $-0.3$ kcal/mol in Ref.\citenum{schmidtEnergeticPositionsLowest1971}.
Then we investigated the diradical character of these two hydrocarbons by analyzing the DCC.
As shown in Ref.\citenum{yangNatureGroundElectronic2016},
the larger the DCC, the less the diradical character.
For the Chichibabin's hydrocarbon,
the configuration with two electrons occupying the HOMO contributes $66.7\%$ to singlet ground state.
For the M\"{u}ller's hydrocarbon,
the configuration with two electrons occupying the HOMO contributes $54.6\%$ to singlet ground state,
which shows a much stronger diradical character. \\

By using the active-space ppRPA approach, 
the above calculations for two large hydrocarbons can be efficiently performed.
For example,
the restricted calculation for the M\"{u}ller's hydrocarbon with the cc-pVDZ basis set contains 147 occupied orbitals and 673 virtual orbitals.
The dimension of the full-space ppRPA matrix for the singlet calculation is 237679.
In the active-space ppRPA approach with 30 active occupied orbitals and 30 active virtual orbitals,
the dimension is reduced to 930. 
The computational cost is greatly reduced compared with the full-space calculation.

\begin{table}
\caption{\label{tab:hydrocarbon}S-T gaps and the dominant configuration contributions of the Chichibabin's hydrocarbon and the M\"{u}ller's hydrocarbon obtained from the active-space ppRPA approaches based on B3LYP. 
30 occupied and 30 virtual frontier orbitals are included in the active space.
Geometries were taken from Ref.\citenum{escayolaEffectExocyclicSubstituents2019}.
Reference values were taken from Ref.\citenum{montgomeryMolecularStructuresThiele1986} and Ref.\citenum{schmidtEnergeticPositionsLowest1971}.
The cc-pVDZ basis set was used.}
\centering
\begin{tabular}{ccccc}
\toprule
                & $\Delta {\text{E}}_{\text{S-T}}$ ref  & $\Delta {\text{E}}_{\text{S-T}}$ calc & HOMO+HOMO & LUMO+LUMO \\
                \midrule
Chichibabin     & -5.5 kcal/mol & -3.0 kcal/mol & 66.9\%    & 28.5\%    \\
M\"{u}ller      & -0.3 kcal/mol & -0.1 kcal/mol & 54.6\%    & 42.5\%    \\
\bottomrule
\end{tabular}
\end{table}

\FloatBarrier

\subsection{CLBEs obtained from active-space T-matrix approaches}

\begin{table}
\setlength\tabcolsep{15pt}
\caption{\label{tab:core65_absolute}MAEs of absolute CLBEs in the CORE65 set obtained from the full-space and the active-space and $G_{0}T_{0}$ and $G_{\text{RS}}T_{\text{RS}}$ approach based on HF, PBE and B3LYP. 
Only occupied orbitals are in the active space. 
The NIP approximation is used to approximate the contribution to the self-energy outside the active space.
Full-space results were taken from Ref.\citenum{liRenormalizedSinglesGreen2021}.
Geometries and reference values were taken from the CORE65 set\cite{golzeAccurateAbsoluteRelative2020}.
The def2-TZVP basis set was used.
All values in eV.}
\centering
\begin{tabular}{cccccc}
    \toprule
                                 &              & HF    & PBE   & B3LYP   \\
    \midrule
    $G_0T_0$                     & full-space   & 3.74  & 14.97 & 9.34  \\
                                 & active-space & 3.81  & 14.58 & 9.43  \\
    $G_{\text{RS}}T_{\text{RS}}$ & full-space   & 3.74  & 1.53  & 1.66  \\
                                 & active-space & 3.81  & 1.58  & 1.68   \\
    \bottomrule
\end{tabular}
\end{table}

\begin{table}
\setlength\tabcolsep{5pt}
\caption{\label{tab:core65_relative}MAEs of relative CLBEs in the CORE65 set obtained from the full-space and the active-space and $G_{0}T_{0}$ and $G_{\text{RS}}T_{\text{RS}}$ approach based on HF, PBE and B3LYP. 
The relative CLBEs are the shifts with respect to a reference molecule, 
$\Delta\text{CLBE}=\text{CLBE}-\text{CLBE}_{\text{ref}}$. 
\ce{CH4}, \ce{NH3}, \ce{H2O} and \ce{CF4} have been used as reference molecules for C1s, N1s, O1s and F1s respectively.
Only occupied orbitals are included in the active space. 
The NIP approximation is used to approximate the contribution to the self-energy outside the active space.
Full-space results were taken from Ref.\citenum{liRenormalizedSinglesGreen2021}.
Geometries and reference values were taken from the CORE65 set\cite{golzeAccurateAbsoluteRelative2020}.
The def2-TZVP basis set was used.
All values in eV.}
\centering
\begin{tabular}{ccccccccccccccc}
\toprule
  &\multicolumn{3}{c} {full-space $G_0T_0$} &\multicolumn{3}{c} {active-space $G_0T_0$} & \multicolumn{3}{c} {full-space $G_{\text{RS}}T_{\text{RS}}$} & \multicolumn{3}{c} {active-space $G_{\text{RS}}T_{\text{RS}}$} \\
  \cmidrule(l{0.5em}r{0.5em}){2-4} \cmidrule(l{0.5em}r{0.5em}){5-7} \cmidrule(l{0.5em}r{0.5em}){8-10} \cmidrule(l{0.5em}r{0.5em}){11-13}  
    & HF   & PBE  & B3LYP& HF   & PBE  & B3LYP & HF   & PBE  & B3LYP & HF   & PBE  & B3LYP \\
  \midrule
  C & 0.33 & 0.85 & 0.41 & 0.27 & 1.37 & 0.82  & 0.33 & 0.39 & 0.46  & 0.27 & 0.32 & 0.45  \\
  N & 0.09 & 1.40 & 0.88 & 0.23 & 1.71 & 1.07  & 0.09 & 0.19 & 0.12  & 0.23 & 0.22 & 0.15  \\
  O & 0.22 & 2.29 & 1.47 & 0.29 & 2.21 & 1.46  & 0.22 & 0.29 & 0.27  & 0.29 & 0.25 & 0.25  \\
  F & 0.13 & 0.22 & 0.11 & 0.14 & 0.22 & 0.13  & 0.13 & 0.22 & 0.07  & 0.14 & 0.21 & 0.24  \\
\bottomrule
\end{tabular}
\end{table}

The performance of the active-space T-matrix approach on predicting QP energies is discussed in this section. 
The active-space $G_{\text{RS}}T_{\text{RS}}$ approach was used to to calculate QP energies.
The $G_{\text{RS}}T_{\text{RS}}$ approach that uses the RS Green's function as the starting point and formulates the effective interaction is shown to predict accurate valence and core QP energies.
As shown in Figure.~1 in the Supporting Information,
the active-space T-matrix approach needs around $50\%$ orbitals in the active space to obtain ionization potentials (IPs) with errors smaller than $0.05$ \,{eV}.
However,
for predicting CLBEs,
the active-space T-matrix approach with the active space consisting of only all occupied orbitals produces errors smaller than $0.1$ \,{eV}, 
which corresponds to a percent error of $0.02\%$.
Thus, 
we focus on predicting CLBEs with the active-space T-matrix approach.

The MAEs of absolute and relative CLBEs obtained from the active-space $G_{\text{RS}}T_{\text{RS}}$ and $G_0T_0$ approach based on HF, PBE, B3LYP are shown in Table.~\ref{tab:core65_absolute} and Table.~\ref{tab:core65_relative}.
For absolute CLBEs,
the active-space $G_0T_0$ approach has errors of around $0.1$ \,{eV} to $0.5$ \,{eV} compared with the full-space results.
The active-space $G_{\text{RS}}T_{\text{RS}}$ approach produces errors smaller than $0.1$ \,{eV} compared with the full-space results.
The active-space $G_{\text{RS}}T_{\text{RS}}$@PBE approach has the smallest MAE of $1.58$ \,{eV} compared with reference values,
which agrees with the full-space results.
For relative CLBEs,
the active-space $G_0T_0$ approach leads to large errors around $0.1$ \,{eV} to $0.5$ \,{eV} compared with the full-space results.
The active-space $G_{\text{RS}}T_{\text{RS}}$ approach provides errors only around $0.03$ \,{eV}.
The active-space $G_{\text{RS}}T_{\text{RS}}$@PBE approach has the smallest MAEs compared with reference values,
which agree with the full-space results.
Therefore,
the active-space $G_{\text{RS}}T_{\text{RS}}$ approach is promising for predicting absolute and relative CLBEs by using an active space consisting of only occupied orbitals and the NIP approximation.

\section{CONCLUSIONS}
In summary,
we developed a new efficient active-space ppRPA approach to predict accurate excitation energies of molecular systems.
In the active-space ppRPA approach,
both indexes in particle and hole pairs are constrained,
which only includes the particle and hole pairs with large contributions to low-lying excitation energies. 
We showed that by using the active space composed of only 30 occupied and 30 virtual orbitals,
the active-space ppRPA approach produces errors smaller than $0.05$ \,{eV} compared with the full-space results for excitations of different characters,
including CT, Rydberg, double, valence and diradicals.
Therefore, 
the scaling of the active-space ppRPA approach is only $\mathcal{O}(N)$,
which is much lower than $\mathcal{O}(N^4)$ in the previous work.
We also combined the active-space ppRPA approach with the NIP approximation in the T-matrix approximation to compute QP energies.
The NIP approximation uses KS orbital energies and orbitals to approximate the contribution to the self-energy outside the self-energy.
The active-space $G_{\text{RS}}T_{\text{RS}}$ approach predicts accurate absolute and relative CLBEs with the MAE around $1.5$ \,{eV} and $0.3$ \,{eV}, respectively.
The active-space formalism is expected to greatly extend the applicability of the ppRPA and the T-matrix approach for large systems.

\begin{acknowledgement}
We acknowledges the support from the National Institutes of Health under award number R01-GM061870 (J.L., J.Y. and Z.C.) and the National Science Foundation (grant no. CHE-2154831) (W.Y.).
\end{acknowledgement}

\section*{SUPPORTING INFORMATION}
See the Supporting Information for the convergence behavior of the active-space T-matrix approach and the numerical results of CLBEs obtained from active-space T-matrix approaches.

\section*{Data Availability Statement}
Data and scripts pertaining to this work have been archived in the Duke Research Data Repository\cite{dukedata}.

\bibliography{ref,software}

\end{document}